\documentclass[aps,amsmath,amssymb,superscriptaddress,footinbib]{revtex4}

\usepackage[english]{babel}
\usepackage{latexsym}
\usepackage{graphics}
\usepackage{subfigure}
\usepackage{epsfig}
\usepackage{amsfonts}
\usepackage{amssymb}
\usepackage{amsmath}
\usepackage{bbm}

\begin{document}

\title{Self-rotating wave approximation via symmetric ordering
of ladder operators}


\author{Jonas Larson}
\affiliation{ICFO-Institute de Ciéncies Fotoniques, Parc Mediterrani de la Tecnologia, E-08860 Castelldefels (Barcelona)}
\author{Hector Moya-Cessa}
\affiliation{INAOE, Coordinaci\'on de Optica, Apdo. Postal 51 y
216, 72000 Puebla, Pue., M\'exico}
\affiliation{Universit\"at Ulm, Abteilung f\"ur Quantenphysik D-89069
Ulm, Germany}
\date{\today}

\begin{abstract}
We show how some Hamiltonians may be approximated using rotating
wave approximation methods. In order to achieve this we use the algebra of boson ladder operators, and transformation formulas between normal and symmetric ordering of the operators. The method presented is studied in two special cases; the Morse and the Mathiue models. The connection with regular perturbation theory is given and the validity of the approximation is discussed.
\end{abstract}

\maketitle

\section{Introduction}\label{sec1}
In this contribution we produce approximations for some
Hamiltonians such as the ones related with the Morse potential and
cosine potential (quantum rotor, or Mathiue differential equation). The approximation is based on
the rotating wave approximation (RWA) \cite{rwamag,allen-eberly}
and symmetric ordering of
ladder operators \cite{mandel}. The RWA is widely used in quantum optical systems and particularly in Jaynes-Cummings types of models \cite{Jaynes,knight} and its extensions \cite{exjc}, for example in ion traps \cite{Wineland}, cavity quantum electrodynamics \cite{cqed} and population transfer in atoms and molecules \cite{stirap}. Also the breakdown of the approximations in similar systems has been considered in several papers \cite{rwabreak}. The RWA is usually used for two (or more) interacting 
subsystems, and in all the above examples, interaction between different subsystems are considered.  When passing to the interaction picture with respect to the free Hamiltonian, one remains with a time
dependent Hamiltonian that under certain conditions on the
frequencies involved may take a simpler form by neglecting terms
that oscillate rapidly \cite{allen-eberly,Jaynes}. A phenomelogical motivation for the approximation is that the neglected terms usually describe non-energy conserving processes, which in the Jaynes-Cummings model discribe simultaneous excitation of the atom/ion and the quantized field. The neglected terms, usually very small within the axperimental parameter regimes, give rise to the Bloch-Siegert shift of the energies \cite{BS}. In this
paper we show that we can also apply a RWA in a single Hamiltonian 
systems.  Developing the potential in a Taylor series and grouping the quadratic term of the Hamiltonian with the kinetic energy to
produce a harmonic oscillator term plus an infinite sum. The
frequency of the {\it artificially produced} harmonic oscillator
(HO) is then used as a reference to apply the RWA to the remaining
sum. We call this HO a self-HO for the Hamiltonian studied. Using the underlying algebra of the oscillator ladder operators and the RWA, closed forms of the infintie sums are given in two special cases. Only Hamiltonain systems containing bound or quasi bound states may be approximated, and the validity depends on the relative "deepth/width" of the potential; the deeper and narrower potential the better approximation. Thus, it is best suited for low excited bound states, which is verified by numerical simulations. It is also shown that the RWA gives the result of regular first order perturbation theory of the self-HO, which again underlines the validity regimes of the approximation. 

We proceed as follows, in the next Section we introduce the self-HO and its Fock states of an arbritary Hamiltonian and derive expressions for the diagonal elements of this Hamiltonian in the Fock basis. This is achieved using the properties of the ladder operators. In sections \ref{sec3} and \ref{sec4}, the connection with the RWA is given and two specific models are considered; the Mathieu and Morse equations respectively.  The validity of the approximation is investigated in Section \ref{numsec} for both studied cases. Secton \ref{sec5} is left for a summery, and in the appendix \ref{app} is is shown how the methods may be used in a more genaral way for calculating various sums.

\section{Diagonal matrix elements in the Fock basis of an arbitrary Hamiltonian}\label{sec2}
Given an arbitrary Hamiltonian we expand the potential term in a Taylor series
(in the following we consider for simplicity the case of unit mass and $\hbar=1$)
\begin{equation}
H=\frac{\hat{p}^2}{2}+{V}(\hat{x})=V^{(0)}(0)+V^{(1)}(0)\hat{x}+\frac{\hat{p}^2}{2}+\frac{\omega^2\hat{x}^2}{2}+\sum_{k=3}^{\infty}\frac{V^{(k)}(0)}{k!}\hat{x}^k
\label{arbitrary}
\end{equation}
with $\omega^2=V^{(2)}(0)$. We let this frequency $\omega$ define the ladder operators \cite{Arfken} for the Hamiltonian systems as 
\begin{equation}\label{lad}
\hat{a}= \sqrt{\frac{\omega }{2}}\hat{x} + i \frac{\hat{p}}{\sqrt{2\omega}} , \qquad
\hat{a}^{\dagger}= \sqrt{\frac{\omega }{2}}\hat{x} - i
\frac{\hat{p}}{\sqrt{2\omega}}
\end{equation}
The eigenstates of the number operator $N=\hat{a}^\dagger\hat{a}$ is the Fock states, labeled $|N\rangle$. The ladder operators incrase or decrease the quantum number $N$ when acted on a Fock state; $\hat{a}^\dagger|N\rangle=\sqrt{N+1}|N+1\rangle$ and $\hat{a}|N\rangle=\sqrt{N}|n-1\rangle$. Thus, only terms of the infinite sum
\begin{equation}
 S=\sum_{k=3}^\infty\frac{V^{(k)}(0)}{2^{k/2}k!}\left(\hat{a}^\dagger+\hat{a}\right)^k.
\end{equation}
of equation (\ref{arbitrary}) that contain an equal number of $\hat{a}^\dagger$:s and $\hat{a}$:s will contribute to the diagonal matrix elements $S_{N}=\langle N|S|N\rangle$. When considering those elemenst we may therefor neglect all other terms of the sum. We have
\begin{equation}\label{rwa1}
\left(\hat{a}^\dagger+\hat{a}\right)^k\Rightarrow\left\{\begin{array}{lll}
\left(\begin{array}{c}k \\ k/2\end{array}\right):(\hat{a}^\dagger
\hat{a})^{k/2}:_W, & & k\,\,\mathrm{even}\\ \\ 0\, & &
k\,\,\mathrm{odd}\end{array}\right.
\end{equation}
where $:(\hat{a}^\dagger \hat{a})^k:_W$ denotes the Weyl (symmetric) ordering of the
operator $\hat{N}^k$. It may be transformed into normal ordering using
\cite{Fuji}
\begin{equation}\label{rwa2}
:(\hat{a}^\dagger
\hat{a})^k:_W=\sum_{l=0}^k\frac{l!}{2^l}\left(\begin{array}{c}k \\
l\end{array}\right)^2\hat{a}^{\dagger k-l}\hat{a}^{k-l}.
\end{equation}
Thus, the new sum $\tilde{S}(\hat{x},\hat{p})$, when the "off-diagonal"
terms have been neglected (this sum will now depend on the operator $\hat{p}$), may be written as
\begin{equation}
\tilde{S}=\sum_{k=0}^\infty\sum_{l=0}^k\frac{V^{(2k)}(0)}{2^k(k!)^2}\frac{l!}{2^l}\left(\begin{array}{c}k\\
l\end{array}\right)^2\hat{a}^{\dagger k-l}\hat{a}^{k-l}.
\end{equation}
The second sum may be taken to infinity (as we may only add zeros)
\begin{equation}\label{sum1}
\tilde{S}=\sum_{l=0}^\infty\sum_{k=0}^\infty\frac{V^{(2k)}(0)}{2^{k+l}(k-l)!^2l!}\hat{a}^{\dagger
k-l}\hat{a}^{k-l}.
\end{equation}
For $k<l$ the above expression is zero and thus we make the substitution
$j=k-l$ and get
\begin{equation}
\tilde{S}=\sum_{l=0}^\infty\frac{1}{2^{2l}l!}\sum_{j=0}^\infty\frac{V^{(2j+2l)}(0)}{2^{j}(j)!^2}\hat{a}^{\dagger
j}\hat{a}^{j}.
\end{equation}
By using the identity
\begin{equation}
\hat{a}^{\dagger j}\hat{a}^j=\frac{\hat{N}!}{(\hat{N}-j)!},
\end{equation}
the sum is diagonalized in the number basis $\{|N\rangle\}$ and we get the diagonal elements of (\ref{arbitrary}) 
\begin{equation}\label{hami0}
\langle N|H|N\rangle=\omega(N+\frac{1}{2})+
\sum_{l=0}^\infty\frac{1}{2^{2l}l!}\sum_{j=0}^\infty\frac{V^{(2j+2l)}(0)}{2^{j}(j)!^2}\frac{N!}{(N-j)!},
\end{equation}
where we have used that $\langle
N|\hat{p}^2/2|N\rangle=\langle N|\omega^2\hat{x}^2/2|N\rangle=\frac{\omega}{2}(N+\frac{1}{2})$. Note that the Fock basis
used to calculate the diagonal elements of the Hamiltonian are
obtained from the quadratic term of the Hamiltonian and we call them $|N\rangle$ instead of the usual notation $|n\rangle$. In the following sections we apply the above to some specific examples.

\section{Cosine potential}\label{sec3}
Using the fact that we can obtain closed forms for the diagonal 
elements of some of the potentials studied in Section \ref{sec2}, and
noting that  the RWA is used to keep constant terms (terms that do
not rotate), in this Section we show that for some Hamiltonians we
can produce approximations under certain circumstances, namely,
when some parameters allow us to perform the RWA.

One very important equation for scientists is the Mathieu equation \cite{math1,stegun,vis} which is identical to the Schr\"odinger equation with a sin or cosin potential, also known as the quantum rotor. Durning the last decade it has gained a new shove of attention due to the growing field of cold atoms in optical lattices \cite{optlat}. Approximate results of the eigenfunctions and eigenvalues of the Mathieu equation have been studied earlier \cite{Portugal,math2,math3}. Known approximation methods from physics, such as the WKB and the Raman-Nath have been applied to the Mathieu equation \cite{math2}. Given the Hamiltonian (reminding that we set $\hbar=1$ and $m=1$)

\begin{equation}
H=\frac{\hat{p}^2}{2}-g_0^2\cos q\hat{x}
\end{equation}
we can expand the cosine as (we neglect the constat term as it
only displaces the energies)
\begin{equation}
H=\frac{\hat{p}^2}{2}+\frac{g_0^2q^2}{2}\hat{x}^2-g_0^2\sum_{k=2}^{\infty}
\frac{(q\hat{x})^{2k}  }{(2k)!}(-1)^k.
\end{equation}
With the above definition of the creation and annihilation operators (\ref{lad}), we get
\begin{equation}
\hat{a}= \sqrt{\frac{g_0 q }{2}}\hat{x} + i
\frac{\hat{p}}{\sqrt{2g_0q}} , \qquad \hat{a}^{\dagger}=\sqrt{\frac{g_0 q }{2}}\hat{x} - i
\frac{\hat{p}}{\sqrt{2g_0 q}},
\end{equation}
and rewrite the Hamiltonian as
\begin{equation}
H=g_0 q(\hat{a}^{\dagger}\hat{a}+\frac{1}{2}) -
g_0^2\sum_{k=2}^{\infty} \left(\frac{q
}{2g_0}\right)^k\frac{(\hat{a}^{\dagger}+\hat{a})^{2k} }{(2k)!}(-1)^k
\end{equation}
In the case $g_0 \gg \frac{q}{4}$ we can do RWA, and thus go to a rotating frame with respect to the free Hamiltonian, $U(t)=\exp\left(-ig_0q\hat{a}^\dagger \hat{a}t\right)$. Using the relations
\begin{equation}
U^\dagger(t)\hat{a}U(t)=\hat{a}\mathrm{e}^{ig_0qt},\hspace{1cm}U^\dagger(t)\hat{a}^\dagger U(t)=\hat{a}^\dagger\mathrm{e}^{-ig_0qt},
\end{equation}
and by only keeping non-rotating tyerms in the Hamiltonian we find
\begin{equation} \label{hami1}
H= g_0 q\left(\hat{N}+\frac{1}{2}\right) -g_0^2\sum_{k=2}\left(\frac{q
}{2g_0}\right)^k\frac{(-1)^k }{(k!)^2}:(\hat{a}^{\dagger}\hat{a})^k:_W
\end{equation}

We now proced as in section \ref{sec2}. Inserting (\ref{rwa2}) into (\ref{hami1}) we have $H=g_0 q\left(\hat{N}-1/2\right) -g_0^2 S$
with
\begin{equation}
S\equiv \sum_{k=2}^{\infty}\sum_{l=0}^k \left(\frac{q
}{2g_0}\right)^k\frac{(-1)^k }{(k!)^2} \frac{l!}{2^l} \left(
\begin{array}{l}
k\\
 m
\end{array}\right)^2 \hat{a}^{\dagger k-l}\hat{a}^{k-l}
\end{equation}
Rearranging terms and like in equation (\ref{sum1}) we can sum
$l$ to infinity to get
\begin{equation}
S= \sum_{l=0}^{\infty}\frac{1}{2^ll!}\sum_{k=2}^{\infty}
\left(\frac{q}{2g_0}\right)^k \frac{(-1)^k }{(k-l)!^2}
  \hat{a}^{\dagger k-l}\hat{a}^{k-l}.
\end{equation}
By expressing
\begin{eqnarray}
\nonumber \sum_{k=0}^{\infty} \left(\frac{q}{2g_0}\right)^k \frac{(-1)^k }{(k-l)!^2}
  \hat{a}^{\dagger k-l}\hat{a}^{k-l}& =& \sum_{k=2}^{\infty} \left(\frac{q
}{2g_0}\right)^k \frac{(-1)^k }{(k-l)!^2}
  \hat{a}^{\dagger k-l}\hat{a}^{k-l}
  \\&+ &1 -\left(\frac{q
}{2g_0}\right) \frac{1}{(1-l)!^2} \hat{a}^{\dagger 1-l}\hat{a}^{1-l},
\end{eqnarray}
we can write $S= S_1 + S_2$ with
\begin{equation}\label{sum2}
S_1= \sum_{l=0}^{\infty} \frac{1}{2^ll!}\sum_{k=0}^{\infty}
\left(\frac{q}{2g_0}\right)^k \frac{(-1)^k }{(k-l)!^2}
  \hat{a}^{\dagger k-m}\hat{a}^{k-l}
\end{equation}
and

\begin{equation}
S_2= \left(\frac{q}{2g_0}\right) (\hat{a}^{\dagger
}\hat{a}+\frac{1}{2})-1.
\end{equation}
Note that, like previous in section \ref{sec2}, the second sum in equation (\ref{sum2}) may be started at $k=l$,
i.e.
\begin{eqnarray} \nonumber
S_1 &=& \sum_{l=0}^{\infty} \frac{1}{2^ll!}\sum_{k=l}^{\infty}
\left(\frac{q}{2g_0}\right)^k \frac{(-1)^k }{(k-l)!^2}
  \hat{a}^{\dagger k-l}\hat{a}^{k-l} \\ &=&
\sum_{l=0}^{\infty} \frac{(-1)^l}{2^ll!}\left(\frac{q}{2g_0}\right)^{l} \sum_{n=0}^{\infty} \left(\frac{q}{2g_0}\right)^{n} \frac{(-1)^n}{n!^2}
  \hat{a}^{\dagger n}\hat{a}^{n}.
\end{eqnarray}
or
\begin{eqnarray}
S_1= e^{-\frac{\beta}{4g_0}}:J_0\left(\frac{q}{2g_0}\hat{a}^{\dagger
}\hat{a}\right):
\end{eqnarray}
or using that $\hat{a}^{\dagger j}\hat{a}^{j}=\frac{\hat{N}!}{(\hat{N}-j)!}$
\begin{eqnarray}
S_1= e^{-\frac{q}{4g_0}}L_{\hat{N}}\left(\frac{q}{2g_0}\right)
\end{eqnarray}
The result obtained here agree with previous results were
approximations to the Mathieu equation are obtained
\cite{Portugal}. The result also gives an interesting relation between zeroth Bessel function and Laguerre polynomials.

Combining the results we find the diagonalized RWA Hamiltonian
\begin{equation}\label{rwacosham}
H=\frac{g_0q}{2}\left(\hat{N}+\frac{1}{2}\right)-g_0^2\mathrm{e}^{-\frac{q}{4g_0}}L_{\hat{N}}\left(\frac{q}{2g_0}\right).
\end{equation}
If we develop the Laguerre polynomials in powers of
$q/g_0$, which should be valid within the regimes of the RWA, and remain to second order we obtain
\begin{equation}
H \approx g_0 q(\hat{N} + \frac{1}{2})
-\frac{q^2}{16}(\hat{N}^2+\hat{N}+\frac{1}{2})-g_0^2.
\end{equation}
It should also be pointed out that the method also works for superimposed cosine potentials; $\sum_kg_k^2\cos(q_kx)$. This has interesting applications in for example solid state physics \cite{super0} and cold atoms in optical superlattices \cite{super}.

We conclude this section by making an analogy betwen the RWA and first order perturbation theory. Given the Hamiltonian
\begin{equation}
H=g_0q\left(\hat{N}+\frac{1}{2}\right)
\end{equation}
with eigenstates $|N\rangle$, we perturb it with 
\begin{equation}
V=-g_0^2\cos(qx)-\frac{g_0^2q^2}{2}x^2.
\end{equation}
The first order corerection to the energy \cite{perturb} becomes
\begin{equation}
\delta E=\langle\Psi_n|V|\Psi_n\rangle=-g_0^2\mathrm{e}^{-\frac{q}{4g_0}}L_{n}\left(\frac{q}{2g_0}\right)-\frac{g_0q}{2}\left(n+\frac{1}{2}\right),
\end{equation}
which regains exactly the same result as the one obtained from the RWA method. This relation between first order perturbation theory and the RWA method is easily shown, using Eq. (\ref{diagrel}), to be valid in any general case.

\section{Morse potential}\label{sec4}
Now let us look at the Morse potential \cite{morse,ll}, from which we get the
Hamiltonian
\begin{eqnarray}
H= \frac{\hat{p}^2}{2} +\lambda^2 (1-\exp[-\alpha (\hat{x}-b)])^2
\end{eqnarray}
This kind of Hamiltonian is commonly used to describe properties of diatomic molecules and other situation with anharmonicity \cite{mol}. The Morse Hamiltonian has turned out to have interesting properties in connection with the WKB approximation \cite{wkb1,wkb2} (the WKB quantization gives the exact spectrum), and with supersymetric quantum mechanics \cite{morsesup}.
 
One can transform the Hamiltonian by means of the displacement
operator $e^{ib\hat{p}}$ so that $H_T=e^{ib\hat{p}}H e^{-ib\hat{p}}=\frac{\hat{p}^2}{2}
+\lambda^2 (1-\exp[-\alpha \hat{x}])^2$. The transformed Hamiltonian is then written in the approximate
form (we neglect odd powers because of RWA, i.e. we are in the
regime $\lambda \gg \alpha$)

\begin{eqnarray}\label{hmorse}
H= \sqrt{2} \lambda\alpha (\hat{N}+1/2) + \lambda^2
 (S_1 + S_2)\label{approx-Morse}\end{eqnarray}
with
\begin{eqnarray}\label{sum}
S_1 = \sum_{k=2} \frac{(2\alpha \hat{x})^{2k}}{(2k)!}, \qquad S_2 =-2
\sum_{k=2} \frac{(\alpha \hat{x})^{2k}}{(2k)!}
 \end{eqnarray}
Following the procedure of the former section we may write
\begin{eqnarray}\label{morses1}
S_1= e^{\frac{\alpha}{\sqrt{2}\lambda}}L_{\hat{N}}(-\frac{\sqrt{2}\alpha}{\lambda})-
\frac{\sqrt{2}\alpha}{\lambda}(\hat{N}+1/2) - 1
\end{eqnarray}
and
\begin{eqnarray}\label{morses2}
S_2 =-2\left(
e^{\frac{\alpha}{4\sqrt{2}\lambda}}L_{\hat{N}}(-\frac{\alpha}{2\sqrt{2}\lambda})-
\frac{\alpha}{2\sqrt{2}\lambda}(\hat{N}+1/2) - 1\right)
 \end{eqnarray}
giving the RWA Hamiltonian
\begin{equation}\label{rwamorseham}
H=\frac{\sqrt{2}\lambda\alpha}{2}\left(\hat{N}+\frac{1}{2}\right)+\lambda^2\mathrm{e}^{\frac{\alpha}{\sqrt{2}\lambda}}L_{\hat{N}}\left(-\frac{\sqrt{2}\alpha}{\lambda}\right)-2\lambda^2\mathrm{e}^{\frac{\alpha}{4\sqrt{2}\lambda}}L_{\hat{N}}\left(-\frac{\alpha}{2\sqrt{2}\lambda}\right)+\lambda^2.
\end{equation}
Using the fact that $\lambda\gg\alpha$ in the RWA validity regime, we expand the Hamiltonian to second order in $\alpha/\lambda$
\begin{equation}\label{appmorse}
H\approx\frac{3}{4}\sqrt{2}\lambda\alpha\left(\hat{N}+\frac{1}{2}\right)+\frac{7\alpha^2}{16}\left(\hat{N}^2+\hat{N}+\frac{1}{2}\right).
\end{equation}

Note that if we write the (transformed) Morse Hamiltonian as
\begin{eqnarray}
H= \sqrt{2} \lambda\alpha (\hat{N}+1/2) + \lambda^2
(1-\exp[-\frac{\alpha(\hat{a}+\hat{a}^{\dagger})}{\sqrt{\sqrt{2}\lambda\alpha}}
])^2 - \frac{2\lambda^2\alpha^2}{2}
\frac{(\hat{a}+\hat{a}^{\dagger})^2}{{2\sqrt{2}\lambda\alpha}}
\label{otro-Morse}
\end{eqnarray}
and express the exponentials in a factorized normal form
\begin{eqnarray}
\exp[\alpha(\hat{a}+\hat{a}^{\dagger})] = e^{\alpha^2/2} \sum_{n=0} \sum_{k=0}
\frac{\alpha^{n+k} \hat{a}^{\dagger n}\hat{a}^k }{n!k!},
\end{eqnarray}
RWA on the above expression keeps only terms $n=k$, i.e. the
double sum becomes a single sum, leading to Laguerre polynomials
of order (operator) $N$. In this form we can recover equation
(\ref{approx-Morse}) via RWA and using equation
(\ref{otro-Morse}).

\section{Validity check of the approximation}\label{numsec}
In this section we numerically study the applicability of the above approximation as funtion of the system parameters.

\subsection{Cosine potential}
\begin{figure}[ht]
\begin{center}
\includegraphics[width=8cm]{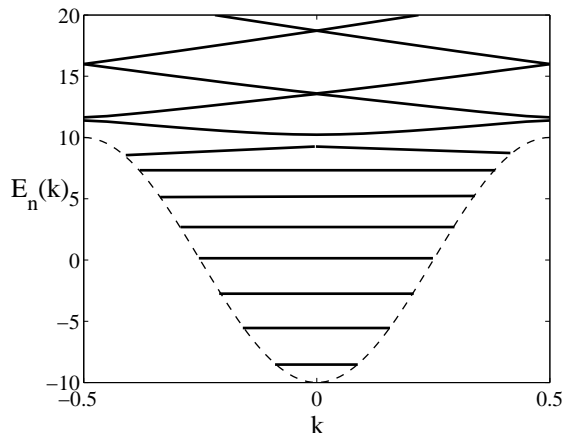}
\caption{\label{fig1} The spectrum $E_n(k)$, within the first Brillouin zone, of the Mathiue equation for a potential amplitude $g_0^2=10$ and wave number $q=1$. The cosine function is plotted for clearity, showing that within the "wells" the particle is quasi bound. }
\end{center}
\end{figure}

As pointed out, solutions of the Mathiue equation in closed analytical form do not exist. However, it is well known, from Floquet \cite{floq} and Bloch theory \cite{bloch}, that the spectrum of a periodic operator in 1-D is determined by a continouos quantum number $k$ refered to as quasi momentum and a descrete number $n$ called band index. The spectrum is most often represented within the first Brillouin zone \cite{bloch} as allowed energy bands seperated by forbidden gaps. The characteristics of the spectrum are usually, in a phenomenological way, explained in two different ways; starting from a weak potential or a strong one. In the weak limit the particle is moving almost freely in a periodic background, which at the degenerate points split the degeneracy. In the oposite, strong potential limit, particles with energies smaller than the potential barriers will be quasi bound, and the tunneling rate determines the width of the bound energies, band width. As soon as a particle has an energy exceeding the barriers, it will move almost freely. Thus, we expect that the $k$-dependence is weak only for the lowest quasi bound energy bands when the amplitudes $g_0$ of the cosine potential is large. This is confirmed in fig. \ref{fig1}, which shows the lowest energy bands of the Mathiue equation for $q=1$ and $g_0^2=10$. 

\begin{figure}[ht]
\begin{center}
\includegraphics[width=8cm]{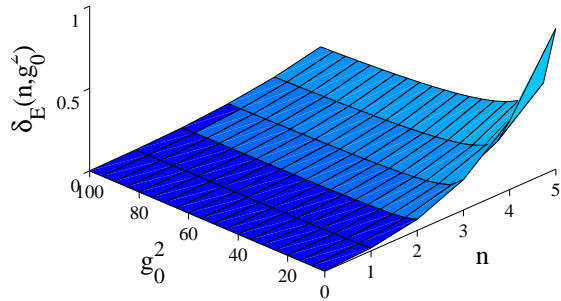}
\caption{\label{fig2} The absolute error $\delta_E(n,g_0^2)=|E_n^{RWA}-E_n|$, between the RWA result and the "exact" numerical result, as a funtion of the band index $n$ and the amplitude $g_0^2$. The approximation is most valid for low excited bands $n$ and strong couplings $g_0$. Again the wave number $q$ is set to unity.}
\end{center}
\end{figure}

The RWA is only supposed to work in the limit of $g_0\gg q$, so it approximate only the lowest quasi bound energies.  In fig. \ref{fig2} we present the results of numerical calculations of the error estimete $\delta_E(n,g_0)=|E_n^{RWA}-E_n|$, where $E_n^{RWA}$ is the approximate result for the energy from eq. (\ref{rwacosham}),
\begin{equation}
E_n^{RWA}=g_0 q\left(n + \frac{1}{2}\right)-g_0^2\left[\mathrm{e}^{-\frac{q}{4g_0}}L_n\left(\frac{q}{2g_0}\right)+\frac{q}{2g_0}\left(n+\frac{1}{2}\right)-1\right]-g_0^2
\end{equation}
and $E_n$ is the numerically calculated result of the energy by diagobnalization of the truncated Hamiltonian. Here the size of the Hamiltonian is $765\times765$ (well within the convergens limits for the eigenvalues). The reason why we show the absolute error, and not the relative one, is because for higher values of $n$, the energies become close to zero and the relative error fluctuates greatly in such cases. In the figure, the band index $n$ runs between 0 and 5, hence showing the six lowest energies, and it is clear that the approximation breaks down for small couplings $g_0$ and high exitations $n$ as expected.

\subsection{Morse potential}
\begin{figure}[ht]
\begin{center}
\includegraphics[width=8cm]{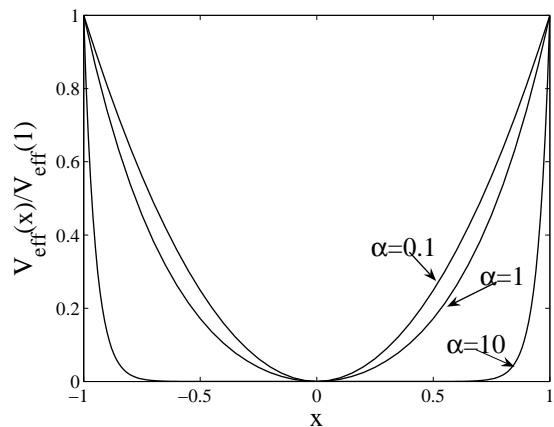}
\caption{\label{fig3} The normalized effective potential $\tilde{V}_{eff}(x)/V_{eff}(1)$ defined in eq. (\ref{apppot}) for $\alpha=0.1,1$ and 10. }
\end{center}
\end{figure}

As the Morse potential is analytically solvable, no numerical diagonalization of the Hamiltonian is needed. The bound energies for the Morse potential are \cite{morse,ll}
\begin{equation}\label{morseexact}
E_n=\sqrt{2}\lambda\alpha\left(n+\frac{1}{2}\right)-\frac{\alpha^2}{2}\left(n+\frac{1}{2}\right)^2.
\end{equation}
Clearly, since $\lambda\gg\alpha$, the second anaharmonicity term becomes crucial only for larger excitations $n$. The anaharmonicity terms is purely negative resulting in that the highly excited bound states are more densly distributed. Interestingly, from the RWA result of eq. (\ref{appmorse}) we note that the second term is positive making the energies to be more sparse for high $n$'s. However, this is partly compensated for by the coeffiecient $3/4$ infront of the first harmonic part. It should be emphasized that only the even terms in the sums (\ref{sum}) are included since within the RWA odd terms vanishes. Thus, it is expected that the method may not be as efficient as for a situation with a purely even potential $V(x)$. It is more likely that the obtained eigenvalues approximate the ones for the Hamiltonian with a potential
\begin{equation}\label{apppot}
\tilde{V}_{eff}(x)=\frac{V_{eff}(x)}{\lambda^2}=1+\sum_{k=0}\frac{(2\alpha x)^{2k}}{(2k)!}-2\sum_{k=0}\frac{(\alpha x)^{2k}}{(2k)!}.
\end{equation}
Using that $\mathrm{e}^{x^2}=\sum_k\frac{x^{2k}}{k!}$, we expect that the potential (\ref{apppot}) has some kind of "weak" exponential behaviour. In fig. \ref{fig3} examples of the normalized effective potential $\tilde{V}_{eff}(x)/V_{eff}(1)$ are given for $\alpha=0.1,1$ and 10.

\begin{figure}[ht]
\begin{center}
\includegraphics[width=8cm]{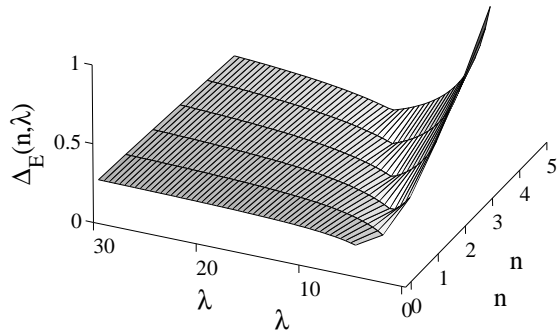}
\caption{\label{fig4} This plot displays the relative error $\Delta_E(n,\lambda)=|E_n^{RWA}-E_n|/E_n$ for $n=0,1,...,5$ and as a function of $\lambda$ and here $\alpha=1$.  }
\end{center}
\end{figure}

The relative error $\Delta_E(n,\lambda)=|E_n^{RWA}-E_n|/E_n$ between the expanded RWA result $E_n^{RWA}$ (\ref{appmorse}) and the exact result $E_n$ (\ref{morseexact}) is shown in fig. \ref{fig4} for the first six eigenvalues and as a function of $\lambda$. In the plot $\alpha=1$, but similar results are obtained for other widths $\alpha$, however, with some increase of $\Delta_E$ for larger $\alpha$.

\section{Conclusions}\label{sec5}
We have developed a method to approximate Hamiltonians using a
kind of self-RWA. We have shown how some of the involved sums may be calculated
using symmetrically ordered expressions for Fock-states
expectation values of powers of the position operator.We have used these expressions to obtain approximations for Hamiltonians
corresponding to the quantum rotor and the Morse potential. A discussion of the validity of the approximation was given and the relation with perturbation theory was explained. The direct link between first order perturbation theory and the RWA deepens the understanding of the two methods, and a consequent question would be if higher orders in the RWA scheeme could regain higher order perturbation theory. This seems possible, but has turned out to be more subtle than expected, mostly because of the non-commutability of the ladder operators. 

\appendix
\section{Calculating sums via symmetric order}\label{app}
In this appendix we use the results of the previous sections and show how one may use it to calculate various sums. Like in section\ref{sec2}, no RWA is used and the results are exact. 

We start with some function $f(x)$ which we can expand according
to
\begin{equation}
f(x)=\sum_{k=0}^\infty\frac{f^{(k)}(0)}{k!}x^k=\sum_{k=0}^\infty\frac{f^{(k)}(0)}{2^{k/2}k!}\left(\hat{a}^\dagger+\hat{a}\right)^k.
\end{equation}
Here we have introduced the ``symmetric'' ladder operators $\hat{a}$ and $\hat{a}^\dagger$ obeying
the regular boson commutator algebra $[\hat{a},\hat{a}^\dagger]=1$ and which
are related to $\hat{x}$ and $\hat{p}=-i\frac{\partial}{\partial x}$ as 
\begin{equation}
\begin{array}{c}
\hat{x}=\frac{1}{\sqrt{2}}\left(\hat{a}^\dagger+\hat{a}\right) \\ \\
\hat{p}=\frac{i}{\sqrt{2}}\left(\hat{a}^\dagger-\hat{a}\right).
\end{array}
\end{equation}
Further we have the normalized eigenstates and eigenvalues of the
number operator $\hat{N}=\hat{a}^\dagger \hat{a}$; $\hat{N}|n\rangle=n|n\rangle$, where
$n=0,1,2,...$, and which in $x$-basis are
\begin{equation}\label{eigfun}
\Psi_n(x)=\langle
x|n\rangle=\frac{\pi^{-1/4}}{\sqrt{2^{n}n!}}\mathrm{e}^{-x^2/2}H_n(x).
\end{equation}

As argued in section \ref{sec2}, the diagonal elements, in the Fock basis, of the function $f(x)$ is identical to the diagonal elements of the function $\tilde{f}(\hat{x},\hat{p})$ where only terms containing an equal number of creation and annihilation operators are included. We thus have 
\begin{equation}\label{diagrel}
\langle n|\tilde{f}|n\rangle=\langle
n|f|n\rangle.
\end{equation}
The L.H.S. can be written, using equation (\ref{hami0}), as
\begin{equation}
 \tilde{f}=\sum_{l=0}^\infty\frac{1}{2^{2l}l!}\sum_{j=0}^\infty\frac{f^{(2j+2l)}(0)}{2^j
(2!)^2}\frac{N!}{(N-j)!}
\end{equation}
and if we express the R.H.S. in the $x$-basis we obtain
\begin{equation}\label{sumintrel}
\sum_{l=0}^\infty\frac{1}{2^{2l}l!}\sum_{j=0}^\infty\frac{V^{(2j+2l)}(0)}{2^{j}(j)!^2}\frac{n!}{(n-j)!}=\frac{1}{2^{n}n!\sqrt{\pi}}\int_{-\infty}^{+\infty}
V(x)\mathrm{e}^{-x^2}H_n^2(x)dx,
\end{equation}
Below we give some analytically solvable examples. Approximate results is in principle easily achievable, but is left out in this paper.

\subsection{Cosine function}
For a cosine function $V(x)=\cos(qx)$ we have
$V^{(2k)}(0)=(-1)^kq^{2k}$ and we find \cite{table} in (\ref{sumintrel})
\begin{equation}
\begin{array}{l}
\mathrm{L.H.S.}=\displaystyle{\sum_{l=0}^\infty\frac{(-1)^lq^{2l}}{4^{l}l!}\sum_{k=0}^\infty\frac{(-1)^kq^{2k}}{2^k(k!)^2}\frac{n!}{(n-k)!}=\mathrm{e}^{-\frac{q^2}{4}}\sum_{k=0}^\infty\frac{(-1)^kq^{2k}}{2^k(k!)^2}\frac{n!}{(n-k)!}}
\\ \\
\mathrm{R.H.S.}=\displaystyle{\frac{1}{2^{n}n!\sqrt{\pi}}\int_{-\infty}^{+\infty}\cos(qx)\mathrm{e}^{-x^2}H_n^2(x)dx=\mathrm{e}^{-\frac{q^2}{4}}L_n(q^2/2)}.
\end{array}
\end{equation}
So that
\begin{equation}
\sum_{k=0}^\infty\frac{(-1)^kq^{2k}}{2^{k}(k!)^2}\frac{n!}{(n-k)!}=L_n(q^2/2).
\end{equation}

Note that closed forms of $V(x)=\exp(\pm iqx)$, $V(x)=\cosh(qx)$,
$V(x)=\cosh^m(qx)$ or $V(x)=\cos^m(qx)$, $m=0,1,2,...$ can also be
obtained.

\subsection{Gaussian function}
In the case of $V(x)=\exp\left(-\alpha^2x^2\right)$, $\alpha^2>0$, the integral
in the R.H.S. of eq (\ref{sumintrel}) is analytically solvable \cite{table}
\begin{equation}
\frac{1}{2^{n}n!\sqrt{\pi}}\int_{-\infty}^{+\infty} \mathrm{e}^{-\alpha^2
x^2}\mathrm{e}^{-x^2}H_n^2(x)dx=2^{n+1}\left(\frac{2\alpha^2}{\alpha^2+2}\right)^{n+1/2}\frac{\alpha^{-1}}{n}F\left(-n,n;-\frac{2n-1}{2};\frac{\alpha^2+2}{2\alpha^2}\right),
\end{equation}
where $F(...,...;...;...)$ is the Gauss hypergeometric function.
The derivatives are simply
$V^{(2k)}(0)=(-1)^k\alpha^{2k}\frac{k!}{(2k)!}$ resulting in the
L.H.S.
\begin{equation}
\mathrm{L.H.S.}=\sum_{l=0}^\infty\frac{(-1)^l\alpha^{2l}}{4^ll!}\sum_{j=0}^\infty\frac{(j+l)!}{(2j+2l)!}\frac{(-1)^j\alpha^{2j}}{2^j(j)!^2}\frac{n!}{(n-j)!}.
\end{equation}
The above holds also for exponential functions such that $0>\alpha^2>-1$, which, when applied to Hamiltonian systems, is of interst as they possess an infinite set of bound states.

\subsection{Hermite polynomials}
When $V_m(x)=H_{2m}(x)$, where $m=0,1,2,...$ we find \cite{table}
\begin{equation}
\mathrm{R.H.S.}=\frac{1}{2^{n}n!\sqrt{\pi}}\int_{-\infty}^{+\infty}
H_{2m}(x)\mathrm{e}^{-x^2}H_n^2(x)dx=\frac{2^{m/2}m!n!}{\left(\frac{m}{2}!\right)^2\left(n-\frac{m}{2}\right)!},
\end{equation}
while its derivatives are
$V_m^{(2k)}(0)=\frac{2^{2k}(2m)!}{(2k)!}H_{2m-2k}(0)$.

\begin{acknowledgments}
J. L. acknowledges Prof. M. Lewenstein, EU-IP Programme SCALA (Contract No. 015714) and the Swedish Goverment/Vetenskapsr\aa det for financial support. H. M.-C. acknowledges E. Kajari for useful discussions and the
Alexander von Humboldt for support. 
\end{acknowledgments}

\end{document}